\documentclass[11pt]{article}

\def\beq{\begin{equation}}
\def\eeq{\end{equation}}
\def\psib{\overline{\psi}}

\usepackage{latexsym}
\usepackage{graphicx}

\begin{document}

\begin{flushright}
\today
\end{flushright}
\bigskip
\bigskip

\begin{center}

{\Large\bf Testing a Fourier Accelerated Hybrid Monte Carlo algorithm.}
\bigskip
\bigskip

{\small Simon Catterall and Sergey Karamov\\
Physics Department, Syracuse University, Syracuse, NY 13244\\}
       
\footnotetext{Corresponding author: Simon Catterall, 
email: smc@physics.syr.edu}

\end{center}

\begin{abstract}

We describe a Fourier accelerated Hybrid Monte Carlo algorithm
suitable for dynamical fermion simulations of non-gauge models. 
We test the algorithm in
supersymmetric quantum mechanics viewed as a one-dimensional
Euclidean lattice field theory. We find dramatic reductions in the
autocorrelation time of the algorithm in comparison to standard HMC.

\end{abstract}

\newpage
\section*{Introduction}

Dynamical fermion algorithms play a crucial role in the simulation of
lattice field theories. The favorite algorithm 
for an even number
of fermion species is the Hybrid Monte Carlo HMC algorithm \cite{hmc}. 
In this paper we introduce and test an improved version of this 
algorithm in the case of
supersymmetric quantum mechanics. We show that
the improved algorithm is
far superior to the usual HMC procedure in combating the
effects of critical
slowing down.

We first discuss the usual HMC algorithm and show how it may
be generalized to allow for Fourier acceleration. The idea is
closely related to the usual Fourier acceleration used for
Langevin simulations - both the fields and their conjugate momenta
are evolved in momentum space throughout an individual HMC
trajectory - the crucial improvement being to choose a wavelength
dependent timestep. The momentum dependence of this
timestep is chosen so as to render the magnitude of
the resulting
update of a given Fourier component insensitive to
its wavevector index.
We test this algorithm in supersymmetric quantum mechanics
treated as a Euclidean lattice theory. This model is formulated
on the lattice in such way as
to leave intact an exact subgroup of the continuum
supersymmetry. We are able to demonstrate that the
improved algorithm reduces the dynamical critical exponent to values 
close to zero on lattices as large as $L=256$ and correlation lengths
$\xi\sim 16$.

\section*{Hybrid Monte Carlo algorithm.}

We will be concerned with simulations of models involving
scalar and fermion fields. The typical action we will
be discussing takes the
general form:
\beq
S(x,\psi,\psib)=S_B(x)+\psib M(x)\psi
\eeq
containing a bosonic field $x$ and a Dirac field $\psi$ 
defined on a lattice in Euclidean space. The fermion
matrix $M$ will contain lattice derivative terms together with
couplings to the bosonic field $x$. The partition function for
this system is then just
\beq
Z=\int DxD\psib\psi e^{-S(x,\psib,\psi)}
\eeq

In order to simulate this action we first replace 
the fermion field by a bosonic pseudofermion field $\phi$ 
whose action is
$$
\sum_{ij}\frac{1}{2}\phi_i\left(M^TM\right)^{-1}_{ij}\phi_j
$$
This is an exact representation of the original boson effective 
action in the case where $\det{M}> 0$ (which will be the case for
SUSY QM). 
The resultant nonlocal action $S\left(x,\phi\right)$ can 
be simulated using the Hybrid Monte Carlo (HMC) algorithm \cite{hmc}. 
In the HMC scheme {\it momentum} fields $(p,\pi)$ conjugate to
$(x,\phi)$ are added and a {\it Hamiltonian} $H$ constructed from
the original action and additional terms depending on the
momenta:
$$
H=S+\Delta S
$$
$$
\Delta S=\sum_i\frac{1}{2}\left(p_i^2+\pi_i^2\right)
$$
The corresponding partition function $Z^\prime =e^{-H}$
is, up to a constant factor, 
identical to the original $Z$. The augmented system $(x,p,\phi,\pi)$
is now associated with a classical dynamics depending on an auxiliary
time variable $t$
$$
\frac{\partial x}{\partial t}=p\ \ \ \ ,\ \ \ \ 
\frac{\partial p}{\partial t}=-\frac{\partial H}{\partial x}
$$
$$
\frac{\partial \phi}{\partial t}=\pi\ \ \ \ ,\ \ \ \ 
\frac{\partial \pi}{\partial t}=-\frac{\partial H}{\partial\phi}
$$

Introducing a finite time step $\Delta t$ allows us to simulate 
this classical evolution and to produce a sequence of configurations 
$(x(t),\phi(t))$. 

If $\Delta t=0$ then $H$ would be conserved along such a trajectory. 
As $\Delta t$ is finite $H$ is not exactly conserved. However a finite
length of such an approximate trajectory can be used as a {\it global} 
move on the fields $(x,\phi)$ which may then be subject to a metropolis 
step based on $\Delta H$. Provided the discrete, classical dynamics is reversible 
and care is taken to ensure ergodicity the resulting move satisfies 
detailed balance and hence this dynamics will provide a simulation of 
$Z^\prime$ and hence also of the original partition function $Z$. 

Ergodicity is taken care of by 
drawing new momenta from a Gaussian distribution after each trajectory.
The reversibility criterion can be satisfied by using a 
leapfrog integration scheme. Its general form for a field $x$
with associated momentum $p$ can be written as \cite{espriu}
\begin{eqnarray}
x_{\bf r}(\delta t)&=&x_{\bf r}(0)+
       \delta t A_{\bf rr^\prime}p_{\bf r^\prime}(0)+
       \frac{(\delta t)^2}{2}A_{\bf rr^{\prime\prime}}
    A^T_{\bf r^{\prime\prime}r^\prime}F_{\bf r^\prime}(0)\nonumber\\
p_{\bf r}(\delta t)&=&p_{\bf r}(0)+
\frac{\delta t}{2}A^T_{\bf rr^\prime}\left(F_{\bf r^\prime}(0)+
                                           F_{\bf r^\prime}(\delta t)\right)
\end{eqnarray}
where $F=-\partial H/\partial x$ is an associated force
and A is an arbitrary matrix. Notice our notation -- the fields are
indexed by a integer vector giving their lattice position.
This update indeed satisfies the time reversibility criterion:
inverting the sign of momentum as $p_n(\delta t)=-p_n(\delta t)$
and updating the resultant configuration leads to 
$\phi_n(2\delta t)=\phi_n(0)$, $p_n(2\delta t)=-p_n(0)$.

To implement our improved algorithm we 
choose a matrix $A_{\bf rr^\prime}$ whose elements depend only on 
the difference of the lattice vectors between two
sites ${\bf r}$ and ${\bf r^\prime}$. 
$A_{\bf rr^\prime}=a\left({\bf r-\bf r^\prime}\right)$ 

We then take the lattice Fourier transform
of equation~{3.}
to arrive at an update equation for
the Fourier components $\hat{x}_{\bf r}$
and $\hat{p}_{\bf r}$. The Fourier components of
the field $x_{\bf r}$ are given by the discrete transform
\beq
\hat{x}_{\bf k}=\frac{1}{\sqrt{V}}\sum_{\bf r} e^{i{\bf k.r}} x_{\bf r}
\eeq
with similar expressions for $\hat{p}_{\bf k}$ and $\hat{a}_{\bf k}$.
Because of the lattice convolution theorem the dependence of
this momentum space update on the Fourier coefficients $\hat{a}_{\bf k}$
may then be completely absorbed into the definition
of a wavelength dependent time step
$\delta t_{\bf k}=\delta t \hat{a}_{\bf k}$.
The final update equations take the form
\begin{eqnarray}
\hat{x}_{\bf k}(t+\delta t_{\bf k})&=&\hat{x}_{\bf k}(t)+
\delta t_{\bf k}\hat{p}_k(t)+
\frac{(\delta t_{\bf k})^2}{2} \hat{F}_{\bf k}(t)\nonumber\\
\hat{p}_{\bf k}(t+\delta t_{\bf k})&=&\hat{p}_{\bf k}(t)+
\frac{\delta t_{\bf k}}{2}\left(\hat{F}_{\bf k}(t)+\hat{F}_{\bf k}(t+\delta t_
{\bf k})\right)
\end{eqnarray}

The final step in applying Fourier acceleration to this update is to
choose the acceleration
kernel $\overline{a}_{\bf k}$ so as 
eliminate the wavelength dependence
of the update in the free theory \cite{bat}. For the update of a 
bosonic field this implies that we should utilize 
the square root of the momentum space lattice propagator. 
Such a propagator contains a mass parameter $m_{\rm acc}$
which can be tuned
to optimize the update. In the free theory it should clearly be
set to the bare lattice mass, but in a general interacting theory
it is left as a parameter. In this paper we provide evidence that
the autocorrelation time is optimized by setting $m_{\rm acc}$ to
the approximate position of the massgap. The fermionic kernel
is then chosen to be the inverse of the bosonic kernel.

Notice the occurrence of
the square root of the momentum space lattice propagator here - rather
than the propagator itself which would be usual in Fourier accelerated
Langevin algorithms - this reflects the fact that here
the HMC update corresponds to 
a discrete second order differential equation in time
unlike the first order Langevin equation. This choice of
the square root propagator was {\it not} made 
in \cite{espriu} and may explain, in part,
why acceleration led to only small reductions in the autocorrelation time.

\section*{Model}
A simple way to demonstrate the effectiveness of
this update is to consider
supersymmetric quantum mechanics \cite{witten}
which contains 
a real scalar field $x$ and two 
independent real fermionic fields $\psi$ and $\psib$ defined on 
a one-dimensional lattice of $L$ sites with periodic boundary 
conditions imposed on both scalar and fermion fields:
\beq
S=\frac{1}{2}\sum_{ij}\left(D_{ij}x_j+P_i\right)
\left(D_{ij}x_j+P_i\right)+
\frac{1}{2}\sum_{ij}\psib_i\left(D_{ij}+P^\prime_{ij}\right)\psi_j
\label{action}
\eeq
For our simulations the quantity $P_i$ and its derivative are chosen as:
$$
P_i=\sum_j K_{ij}x_j+gx_i^3
$$
$$
P^\prime_{ij}=K_{ij}+3gx_i^2\delta_{ij}
$$
The symmetric difference operator $D_{ij}$
and the Wilson mass matrix $K_{ij}$ are defined as:
$$
D_{ij}=\frac{1}{2}\left[\delta_{j,i+1}-\delta_{j,i-1}\right]
$$
$$
K_{ij}=m\delta_{ij}-\frac{1}{2}\left(\delta_{i,j+1}+
\delta_{i,j-1}-2\delta_{ij}\right)
$$
Here dimensionless lattice units are $m=m_{\rm phys}a$, 
$g=g_{\rm phys}a^2$ and $x=a^{-\frac{1}{2}}x_{\rm phys}$
and the discrete momenta take on the values $2\pi k/L$ with
$k=0\ldots L-1$.
Notice that this model employs a non-standard boson action containing
not the usual scalar lattice Laplacian but the square of the
symmetric difference operator. This is done in order to treat
the fermions and bosons in a symmetric manner - indeed because
of this the action
is invariant under a single SUSY-like symmetry.
\begin{eqnarray}
\delta x_i &=& \psi_i\xi \nonumber\\
\delta \psib_i &=& (D_{ij}x_j+P_i)\xi \nonumber \\
\delta \psi_i &=& 0 \nonumber
\end{eqnarray}
Doubles in both
bosonic and fermionic sectors are eliminated by means of the
Wilson mass term $K$. The physics results from this study were
published in \cite{us1}.

For bosonic and pseudo-fermionic field updates respectively we use
the following timesteps which are simple inverses of each other.
$$
\delta t_k^B=\Delta t (m_{acc}+2)/
\sqrt{\sin^2 (2\pi k/L)+(m_{acc}+2\sin^2(\pi k/L))^2}
$$
$$
\delta t_k^F=\Delta t 
\sqrt{\sin^2 (2\pi k/L)+(m_{acc}+2\sin^2(\pi k/L))^2}/(m_{acc}+2)
$$

\section*{Autocorrelation time}

Suppose that $Q$ is some observable and
$Q_t$ is a measurement of $Q$ in configuration
corresponding to Monte-Carlo time $t$. Then the
autocorrelation function is defined as:
$$
c(t)=\frac{<Q_0Q_t>-<Q>^2}{<Q^2>-<Q>^2}
$$
Typically the function $c(t)$ is approximately exponential $c(t)=e^{-t/\tau}$
where $\tau$ is the autocorrelation time - the time between 
decorrelated configurations. Clearly we can define $\tau$ also
from the relation 
$$
\tau=\int_0^\infty e^{-t/\tau}dt=\sum_t c(t)
$$
and this yields a robust way to estimate $\tau$ even when $c(t)$ is
not exactly an exponential - measured this way it is
sometimes referred to as the integrated autocorrelation time. It also
provides
a convenient way to find the autocorrelation time from a sequence of $N$
Monte Carlo measurements
\beq
\tau=\sum_{i=1}^N\sum_{j=i}^{j<M}\frac{Q(i)Q(j)-<Q>^2}{<Q^2>-<Q>^2}
\eeq
where we measure the autocorrelation function only out to $M$ steps
(we must choose $M>>\tau$).

The efficiency of different simulation algorithms
can be described in terms of the dynamic critical exponent $z$:
$$
\tau \sim (1/M_{\rm lattice})^z=(L/M_{\rm physical})^z\sim L^z
$$
An algorithm which resulted in $z=0$ is said to suffer no
critical slowing down while most 
local algorithms such as conventional HMC typically yield $z\sim 2$.

\begin{figure}[htb]
\begin{center}
\includegraphics[width=11cm]{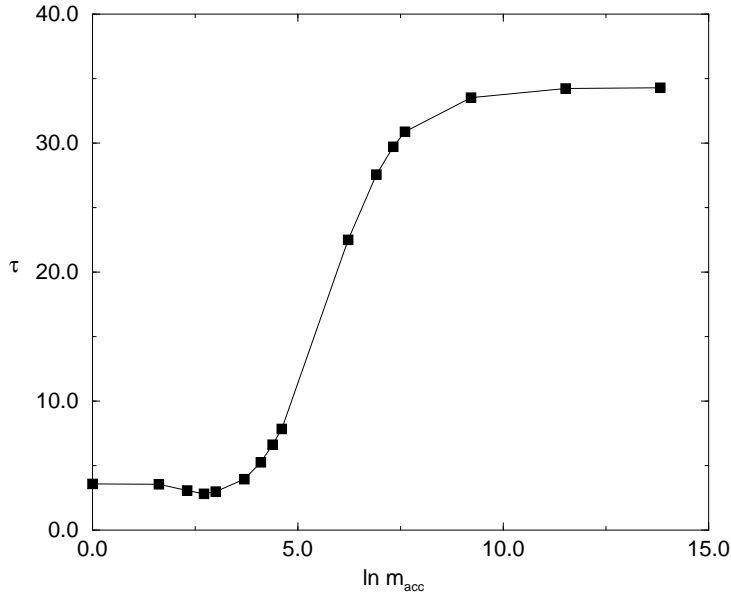}
\end{center}
\caption{\label{fig1} $\ln{\tau}$ vs $\ln{m_{acc}}$ 
for
Supersymmetric Quantum Mechanics with $L=64$, $m=10$, $g=100$.}
\end{figure}

Fig.~\ref{fig1} represents the autocorrelation time $\tau$
as a function of $m_{\rm acc}$
for lattice SUSY QM with 
$m=10$, $g=100$ for a lattice with $L=64$ sites. 
Notice that the dimensionless parameter
characterizing the interaction strength $g/m^2$ is unity
in this case - we are in a strong coupling regime. The massgap
of the theory for these values of the bare parameters can be calculated
via Hamiltonian methods and yields a value $m_g\sim 16.0$.

For large $m_{\rm acc}\to\infty$ the update can be shown to
reduce to the usual local HMC algorithm and, as the plot makes
clear, leads to
a maximal value for $\tau$. Indeed the minimal value of $\tau$ is
can be seen to be achieved by setting 
$m_{\rm acc}$ approximately equal to the massgap of the model
$\ln{m_{acc}}\sim 2.7$.  
In the vicinity of this point,
one can see that the minimum autocorrelation
$\tau_{\rm min}$ is more than an order of
magnitude smaller than the $\tau$ achieved by the usual local
algorithm 
for this lattice size.

\begin{figure}[htb]
\begin{center}
\includegraphics[width=11cm]{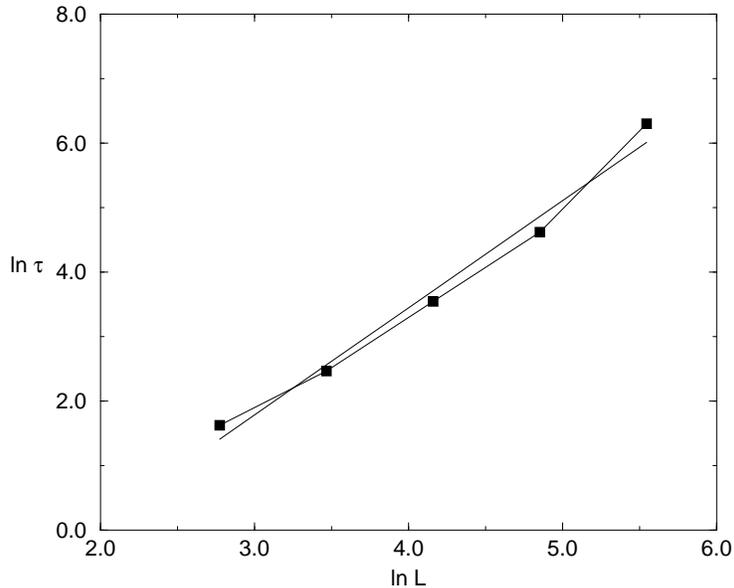}
\end{center}
\caption{\label{fig2} $\ln{\tau}$ vs  $\ln{L}$ 
for
Supersymmetric Quantum Mechanics with $m_{acc}\sim \infty$, 
$m=10$, $g=100$. Straight line fit yields $z=1.7$}
\end{figure}

\begin{figure}[htb]
\begin{center}
\includegraphics[width=11cm]{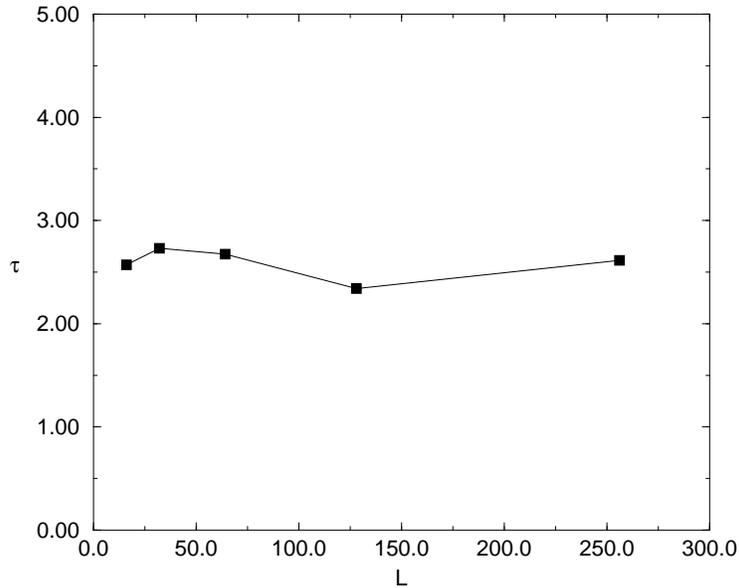}
\end{center}
\caption{\label{fig3} $\ln{\tau}$ vs $\ln{L}$ for
Supersymmetric Quantum Mechanics with $m_{acc}=15$, $m=10$, $g=100$.}
\end{figure}

To derive a dynamic critical exponent we have also compared the
dedpendence of the autocorrelation time on lattice size $L$ 
both for the usual local HMC and the Fourier
accelerated algorithm with $m{\rm acc}=15.0$.
For the case $m_{\rm acc}=\infty$ fig~\ref{fig2} shows
a plot of $\ln{\tau}$ against $\ln{L}$ together with
a straight line fit yielding $z\sim 1.7$. The plot
makes it clear that this is a lower bound for $z$ 
so that in all likelihood $z$ achieves a value of $z=2$
for very large lattices.
This is
to be contrasted to fig~\ref{fig3} which shows the same plot for
$m_{\rm acc}=15$. A fit in this case yields a value of $z$
consistent with zero. For the largest lattice size $L=256$ we see
that more than two orders of magnitude decrease
in $\tau$ are gained by use of the
accelerated algorithm.

\section*{Discussion}

We have shown that the use of Fourier acceleration can yield large payoffs
in the simulation of non-gauge field theories with dynamical
fermions. We have presented results which support this conclusion
for supersymmetric quantum mechanics in which the HMC algorithms can
be pushed to large ($L=256$) lattice size and correlation
length ($\xi\sim 16$). We have obtained similar, though less
quantitative, results for two-dimensional Wess-Zumino models
\cite{wz}.

\section*{Acknowledgements}
Simon Catterall was supported in part by  
DOE grant DE-FG02-85ER40237

\end{document}